\documentclass[useAMS,usenatbib]{mn2e}
\def\simlt{\lower.5ex\hbox{$\; \buildrel < \over \sim \;$}}
\def\simgt{\lower.5ex\hbox{$\; \buildrel > \over \sim \;$}}
\def\ea{et al.\ }
\def\cm3{{\rm cm^{-3}}}
\def\kms{km s$^{-1}$}
\def\schi{{\sc Hi}\ }
\def\msol{{$M_\odot$}}

\def\tmerge{t_{\rm merge}}
\def\aeff{A_{\rm eff}}
\def\vlsr{v_{\rm LSR}}

\def\vmin{v_{\rm min}}
\def\vmax{v_{\rm max}}
\def\vcr{v_{\rm cr}}

\def\dmax{d_{\rm max}}
\def\tsf{t_{\rm sf}}
\def\rsf{R_{\rm sf}}
\def\vsf{v_{\rm sf}}
\def\msf{M_{\rm sf}}
\def\dsf{d_{\rm sf}}
\def\tpds{t_{\rm pds}}

\def\sigmasn{\sigma_{\rm SN}}
\def\pgb{p_{\rm B}}
\def\ptn{p_{\rm T}}
\def\fhi{f_{\rm HI}}
\def\fobs{f_{\rm obs}}
\def\fhalo{f_{\rm halo}}
\def\fhole{f_{\rm hole}}
\def\lcl{\lambda_{\rm cl}}
\def\acl{a_{\rm cl}}
\def\fcl{f_{\rm cl}}
\def\ncl{n_{\rm cl}}
\def\nrad{N_{\rm rad}}
\title [Visibility of Old Supernova Remnants in \schi 21-cm Emission Line]
{Visibility of Old Supernova Remnants in \schi 21-cm Emission Line}
\author [B.-C. Koo and J.-h. Kang] {Bon-Chul Koo\thanks{E-mail: koo@astrohi.snu.ac.kr 
(BCK); kjh@astro.snu.ac.kr (JhK)} and Ji-hyun Kang\footnotemark[1] \\
Astronomy Program, SEES, Seoul National University, Seoul 151-742, Korea}
\begin{document}

\date{Accepted 2003. Received 2003; in original form 2003 August}

\pagerange{\pageref{firstpage}--\pageref{lastpage}} \pubyear{2003}

\maketitle

\label{firstpage}

\begin{abstract}

We estimate the number of old, radiative supernova remnants (SNRs) {\em detectable} in
\schi 21-cm emission line in the Galaxy. We assume that old SNRs consist of
expanding \schi shells and that they are visible if the line-of-sight velocities
are sufficiently outside the velocity range of the Galactic background \schi emission.
This criterion of visibility makes it possible to calculate the
background contamination and to make a comparison with observation.
The Galactic disk in our model is filled with atomic gas of 
moderate ($\sim 0.1$~cm$^{-3}$) density representing the warm neutral 
interstellar medium.
We assume that only Type Ia supernovae produce isolated
SNRs with expanding \schi shells, or ``\schi SNRs". 
According to our result, the contamination
due to the Galactic background \schi emission limits
the number of visible SNRs to $\simeq 270$, or $\simeq 9$\% of the total \schi SNRs.
They are concentrated along the loci of tangential points.
The telescope sensitivity further limits the number.
We compare the result with observations to find that the observed
number ($\le 25$) of \schi SNRs is much less than the expected.
A plausible explanation is that previous
observational studies, which were made toward the SNRs identified mostly in radio continuum,
missed most of the \schi SNRs because they are too faint to be visible in radio continuum.
We propose that the faint, extended \schi 21-cm emission line 
wings protruding from the Galactic background \schi emission
in large-scale $(\ell,v)$ diagrams could be possible candidates
for \schi SNRs, although our preliminary result shows that
their number is considerably less than the expected in the inner Galaxy.
We conclude that
a possible explanation for the small number of \schi SNRs in the
inner Galaxy is that the interstellar space
there is largely filled with a very tenuous gas as in the
three-phase ISM model, not with the warm neutral medium of moderate density.

\end{abstract}


\section{Introduction}

A general picture of old SNRs is a hot bubble surrounded by
a dense, atomic shell. The shell is preceded by SNR shock, which is radiative.
A shock is referred to as ``radiative" if the cooling time of the shock-heated
material is shorter than the age of the shock. For the SNR shock propagating in a
uniform, homogeneous medium of hydrogen density $\sim 1$~cm$^{-3}$, it becomes radiative
when the SNR is $\sim 3.6\times 10^4$~yrs old, or 
when the shock velocity drops to $\sim 180$~\kms\ (see Section 2.3). 
The shell decelerates as it sweeps out more material, and, eventually, when the SNR is
$\sim 10^6$~yrs old, the
velocity of the shell drops to an order of the ambient sound speed ($\sim 10$~\kms) and
the SNR merges into the general interstellar medium (ISM). 
We may call old SNRs with fast-expanding \schi shells ``\schi SNRs" because they 
are observable in \schi 21-cm emission line.

A systematic \schi 21-cm line observation of SNRs to detect
the expanding shells have been made by Koo \& Heiles (1991; hereafter KH91).
It is difficult to detect expanding \schi shells,
because most known SNRs are located in the Galactic plane where the
Galactic background \schi emission causes severe contamination.
There had been papers reporting the detection of \schi shells
with expansion velocities smaller than $\sim$20~\kms,  
but the association was rather ambiguous
\citep{ass73, kna74, col80, col82}.
%In some SNRs, high-resolution studies revealed \schi
%features around the systematic velocity of SNRs which are
%spatially correlated with the radio continuum
%filaments of the SNRs (e.g., Braun \& Strom 1986)
%but such high-resolution studies are limited.
If the expansion velocity of the shell is very large, larger than
the maximum velocity permitted by the Galactic rotation, however, then it
could be easily discernable from the background emission.
Fig. ~1 shows such an example, where we can clearly see an
extended high-velocity (HV) excess emission
localized at the position of the SNR HB~21.
The positional coincidence and the forbidden velocity
strongly suggest that the excess emission
is emitted from the gas accelerated by the SNR shock.
KH91 carried out a sensitive survey
of \schi 21~cm emission lines toward
103 northern Galactic SNRs and detected such HV gas
%`forbidden-velocity (FV) emission line wings'
toward 15 SNRs\footnote{KH91 observed each SNR at 9 points in a cross
pattern centered at its catalog position. They searched for excess emission wider than
10~\kms\ and divided the SNRs into 3 ranks, in which increasing number implies
increasing reliability of the detected \schi feature. These 15 SNRs include
the fourteen rank 3 SNRs (excluding G117.4+5.0 which is not considered as a 
SNR any more [Green 2001])
%for which KH91 concluded that the associated HV
%\schi gas is quite likely due to
%the gas accelerated by the SNR shock 
and the classical source IC~443. 
}, including 3--4 SNRs known prior to the survey
(DeNoyer 1978; Giovanelli \& Haynes 1979;
Landecker, Roger, \& Higgs 1980; Braun \& Strom 1986;  Koo \ea 1990).
%\citep{den78, gio79, lan80, bra86, koo90}.
Koo, Kang, \& McClure-Griffiths (2003) searched for similar \schi features
%\cite{koo03b} searched for similar \schi features
toward 97 southern SNRs
using the Southern Galactic Plane Survey data (McClure-Griffiths 2001)
and identified another 10 SNRs. 
These $25 (=15+10$) SNRs are {\em candidates} for \schi SNRs because 
the surveys were made
using single-dish radio telescopes with a large ($36'$ or 16$'$) beam size and 
high-resolution observational studies are necessary
for confirming the nature of the HV emission.
Such high-resolution studies have been done for some of the SNRs and confirmed
the shell-like structure of the HV \schi features, e.g.,
CTB 80 \citep{koo90}, W44 \citep{koo95}, W51C \citep{koo97},
and IC 443 \citep{gio79, bra86}.
But molecular lines studies showed that the latter three 
SNRs exhibit evidences for
the interaction with molecular clouds, e.g., broad molecular lines,
high-ratio of high- to low-transition CO lines, 1720-MHz maser line
of the OH molecule, etc. (see Koo 2003 and references therein).
The HV \schi gas in these SNRs
is possibly formed by a radiative shock propagating through
molecular clouds, not by the shock propagating through the general diffuse ISM.
In summary, including the ones interacting with molecular clouds,
there are 25 \schi SNR candidates which appear to have shock-accelerated \schi gas at velocities
forbidden by the Galactic rotation. 

How does this compare to the expected number of \schi SNRs in the Galaxy?
If we naively multiply the total supernova rate ($\sim 0.02 $~yr$^{-1}$)
in the Galaxy to the lifetime ($\sim 1\times 10^6$~yr) of SNRs, we get
$\sim 2\times 10^4$.
The majority of supernovae (SNe), however, are core-collapse SNe with massive
progenitors. Massive stars are born in associations, and most of these SNe
dissipate their energy in producing superbubbles instead of
isolated SNRs. Also, a significant fraction of SNe occur in
the Galactic halo where they merge into the ambient ISM before cooling becomes important.
And, if the structure of the ISM is close to the
three-phase model of the \cite{mck77} so that
most of the interstellar space is filled with
hot tenuous ($3.5\times 10^{-3}$~cm$^{-3}$) medium, 
the SNRs may overlap with other SNRs before dense shell formation.
Meanwhile, the majority of \schi SNRs may not be visible 
because of the contamination due to the Galactic background emission and
the limited sensitivity. The effect of the background contamination is 
difficult to calculate in general, which hampers the estimation of the number of 
visible SNRs. But, for the \schi SNRs, if we limit to the ones with large 
expansion velocities as in Fig. 1, it is straightforward to calculate 
the effect because the boundary of the background emission in Fig. 1 is 
mainly determined by the Galactic rotation.
Therefore, the statistics of such \schi SNRs can be related to 
the Galactic distribution of SNe, their environments, and the evolution 
of their remnants.

In this paper, we develop a model
to estimate the visibility of \schi SNRs in the Galaxy. 
The outline of the paper is as follows: in Section 2 we describe 
our model and formulation. We model the Galactic disk as 
an axi-symmetric, uniform, homogeneous disk with a central hole.
We estimate the frequency and distribution of SNe in the disk using recent
observational results on extragalactic SNe and old disk stars. We summarize
the dynamical evolution of old SNRs in homogeneous, uniform medium and
devise the detection probability. Section 3 presents
the results, where we show the expected distribution of \schi SNRs in the
Galactic plane and their statistics.
We explore how the background contamination and the
telescope sensitivity limit the visible number of \schi SNRs. In Section 4, we
discuss the effect of embedded clouds and 
compare the result with observations.

\section {DESCRIPTION OF MODEL}
\subsection{Galactic \schi Disk and \schi SNRs}

What would be the nature of the ISM that a typical isolated SNR experiences?
This is perhaps the very essential question in deriving the number of \schi SNRs
because the formation and evolution of dense radiative shells
depend on the physical properties of the ambient ISM.
A general picture of the ISM is dense clouds immersed in a
diffuse intercloud medium, perhaps except regions surrounding
OB stars where the clouds might have vanished through photoevaporation and rocket effect
due to strong UV radiation (e.g., McKee, Van Buren, \& Lazareff 1984).
%\citep[e.g.,][]{mck84}.
The birthplaces for the isolated SNe are presumably
far from OB stars, so as to be close to the general ISM.
Then, although the embedded clouds may modify the internal structure
through evaporation,
it is the intercloud medium that primarily
controls the dynamical evolution of SNRs (e.g., Cowie, McKee, \& Ostriker 1981),
%\citep{cow81},
and the question is what the physical conditions of the intercloud medium are.
There have been substantial studies
on this subject, and it is generally considered that the major
component of the intercloud medium is either the
warm neutral medium (WNM) with a typical density of $\sim 0.1$~cm$^{-3}$
as in the two-phase model of Field, Goldsmith, \& Habing (1969)
%\cite{fie69}
or the hot ionized medium (HIM) of $3.5\times 10^{-3}$~cm$^{-3}$
as in the three-phase model of \cite{mck77} (see McKee 1995 for a review).
%Warm ionized medium may also
%occupy non-negligible interstellar space.
Relative filling factors of the two phases, however, are poorly known even in 
the solar neighborhood.
If it is the HIM, then most of the SNRs might dissipate
their energy without forming a radiative shell, and there will be only
a few \schi SNRs. 
In the following, we calculate the number of \schi SNRs assuming that 
the intercloud medium is composed with the WNM only, and, 
by comparing the result with 
observations, we will infer the structure of the ISM. 

We assume that the \schi gas in the Galaxy is confined to
an axi-symmetric, center-emptied disk with an inner edge at 3.5 kpc and outer edge at 15 kpc.
The total \schi column density perpendicular to the Galactic plane is
approximately constant between 3.5 kpc and 20 kpc, but has a central hole.
The inner edge represents the boundary where the \schi surface density drops
rapidly \citep{kul87}.
We take the outer edge of the disk at 15 kpc (instead of 20 kpc) considering that
the distribution of isolated SNe are
possibly truncated at $\simlt 15$~kpc (see Section 2.2). Even if there is no truncation,
the small disk size should not notably affect the result because, as we discuss
in Section 2.2, the isolated SNe are considered to be
centrally concentrated with a short ($\sim 3$ kpc)
exponential radial scale length.
We assume that the thickness of the disk is constant and the
density distribution is uniform.
The real Galactic disk is warped and flares outside the solar circle
(Kulkarni, Blitz, \& Heiles 1982; Lockman 1984). But this complexity should have little
effect on the result because, firstly, most ($\sim 80$\%) SNe occur within the solar circle and, 
secondly, the stellar disk and, therefore the SNe disk, warps and flares too \citep{lop02}.
Our formulation is essentially two-dimensional and the primary purpose of considering 
the thickness of the disk is to
estimate the fraction of SNe that explodes in the very tenuous
halo, which may not evolve to \schi SNRs (Section 2.2).
According to KH91, all the \schi SNR candidates in the
inner Galaxy is within $z<200$~pc, while, in the outer Galaxy, the one (CTA 1) located
at the farthest from the midplane is at $(r,z)=(10~{\rm kpc}, 370~{\rm pc})$
where $r$ is the Galactocentric radius and $z$ is the height above the midplane. The
latter corresponds to $314$~pc at $r=r_\odot\equiv 8.5$~kpc
if we consider that the thickness of the disk flares almost linearly outside the solar circle.
Considering this, we adopt the half-thickness of our disk to be 320 pc, i.e.,
SNe occurring outside this disk are assumed not to form an \schi shell.
This is greater than the half-width at half-maximum (265 pc) of 
the thick Gaussian component and less than the scale height (403 pc) of the 
exponential component of the \schi disk \citep{dic90}.
The total intercloud \schi column density perpendicular to the Galactic plane in the solar
neighborhood is $\sim 90$~cm$^{-3}$ pc \citep{hei87}, and we require our model disk to
have the same vertical column density. Therefore,
the density of our disk $n_0 = 0.14~{\rm cm}^{-3}$.

For such axi-symmetric,
two-dimensional disk, the number of observable \schi SNRs can be written as
\begin{equation}
N= (\Sigma_{\rm SN} \fhi) \tau \fobs, 
\end{equation}
where $\Sigma_{\rm SN}$ is the total {\it isolated}
SN rate in the Galaxy, $\fhi$ is the  fraction of isolated SNe that produces
\schi SNRs (i.e., the fraction within the center-emptied model gaseous disk),
$\tau$ is the lifetime of \schi SNRs, and
$\fobs$ the fraction of \schi SNRs that are detectable:
\begin{equation}
\fobs \equiv {1 \over \Sigma_{\rm SN} \fhi \tau}
\int\,  \sigmasn(r) \int \, p(x,y,t) dt dA  
\end{equation}
where $\sigmasn (r)$ is the surface density distribution of SNe, i.e.,
the SN rate per unit area as a function of Galactocentric radius $r$,
and $p(x,y,t)$ is the probability that a \schi SNR of age $t$
is detectable considering the background contamination and the telescope sensitivity.
$(x,y)$ represents a coordinate grid centered at the Galactic center and
the integration range over $(x,y)$ is from $r=3.5$~kpc to 15 kpc.
If the detection probability
$p(x,y,t)$ is independent of $(x,y)$ and is equal to 1 during $\tau$,
then $\fobs=1$ and $N=\Sigma_{\rm SN} \fhi \tau$.
In the following sections, we estimate the individual items in equations (1) and (2).

\subsection{Frequency and Distribution of SNe}

For our purpose, SNe are divided into two types: Supernovae of type Ia (SNe Ia)
which have old disk stars
as their progenitors and the ``core-collapse" SNe (e.g., type Ib/c and type II SNe;
hereafter SNe II) which have massive progenitors.
SNe II may be further divided into two groups; ones
exploding in associations and the other exploding isolated.
Most of the former SNe are expected to
explode within a superbubble where the
density is very low and pressure is very high. Such SNRs might
dissipate their energy before radiative cooling becomes important, and we
consider that they do not evolve to \schi SNRs.
The latter group is consisted of ``runaway" OB stars which are ejected from their parent
association and also possibly the OB stars formed isolated.
\cite{gie86} derived that
10--25\% of O stars are runaways while only about 2\% of early B stars, which contribute
most of the SNe, are runaways. This led \cite{mck95} to conclude that there is
no need to assume a separate population of isolated OB stars.
On the other hand,
\cite{fer95} analyzed the catalog of OB stars compiled by
\cite{gie87} and \cite{hum84}, and
concluded that 40\% of SNe II occurs isolated (see also McKee 1990).
Among the historical SNe II, at
least Crab  is not associated with a group of OB stars and its
progenitor could have been a runaway \citep{mdz01}.
We assume that all SNe II explode in groups and that only SNe Ia
produce isolated SNRs that can evolve to \schi SNRs. The basic idea is 
to choose parameters to minimize the number of \schi SNRs when they are uncertain, 
so that the derived number is to be a lower limit. 
If the population of isolated OB stars
is indeed significant, then the expected number of
\schi SNRs will be larger than the calculated below.

The Galactic SN Ia rate is inferred from the extragalactic SN rates,
which are usually expressed in SNu or SN (100 yr)$^{-1}$ ($10^{10} L_B/L_\odot)^{-1}$
where $L_B$ is the blue luminosity of the galaxy
%\citep[e.g.,][]{van94, cap99}.
(e.g., van den Bergh \& McClure 1994; Cappellaro, Evans, \& Turatto 1999).
Our Galaxy is considered to be a spiral of Hubble type
Sb or Sbc with $L_B = (2.3\pm 0.6) \times 10^{10}$~$L_\odot$
\citep{van88}, and we take the arithmetic mean of the
average rates of S0a-Sb and Sbc-Sd galaxies in \cite{cap99} to obtain
the total SN Ia rate of $0.45\times 10^{-2}$ yr$^{-1}$. (For
comparison, the total SN II rate is $1.8\times 10^{-2}$~yr$^{-1}$.)
It is noted that
this SN Ia rate is higher than the one (0.3--0.4$\times 10^{-2}$)
derived from the nova rate \citep{van91}, but
consistent with the estimate (0.3--0.6$\times 10^{-2}$) of \citet{van94}.

SNe are centrally concentrated and vertically stratified, presumably
following the distribution of their progenitors.
The progenitors of SN Ia are mostly old disk stars.
Bulge stars could contribute as much as $15$\% to the total SN Ia rate
if the SN rate scales with the mass ratio of the bulge ($\sim 10^{10}$~\msol) to the
disk ($\sim 7\times 10^{10}$~\msol) \citep{daw94}. But the rate of SN Ia within 1 kpc of
the center of spiral galaxies has been found to be significantly lower than
outside this region
(Wang, H\"oflich, \& Wheeler 1997; Hatano, Branch, \& Deaton 1998), which suggests
that the SN Ia occurs in the bulge not as frequently as expected.
The space density of old disk stars perpendicular to the Galactic plane can be
accurately represented by the sum of two exponentials, one
with a characteristic scale height of $\sim 300$~pc and the other
which has a scale height of $\sim 1$~kpc \citep[e.g.,][]{bin98}.
The latter ``thick disk" stars appear older ($\simgt 10^{10}$~yr)
and more metal poor ([Fe/H]$\simlt -0.4$) than the former ``thin" disk
\citep{bin98}. The total mass of the thick disk stars
is $\simlt 10\%$ of the thin disk stars (Trimble 2000). Therefore, if
the SN Ia rates of thin and thick disks are proportional to their masses,
the thick disk could contribute at most $\simlt 10$\%
of the thin disk to the total SN Ia rate.
We neglect the small uncertain contributions from the bulge and thick disk,
and assume that the old stellar disk representing the SN Ia progenitors has
a vertical exponential distribution with a scale height of 300 pc at $r=r_\odot$.
This implies that, in the solar neighborhood, 34\% of SNe Ia explodes outside
our $\pm 320$~pc-thick gaseous disk.
We further assume that this fraction remains constant over the full disk, e.g.,
we assume that the fraction of SNe Ia exploding in the `\schi halo' and
not evolving to \schi SNRs $\fhalo=0.34$. 
This
is not unreasonable because the old stellar disk warps and flares as the gaseous
disk does \citep{fre98, lop02}.
But the flaring of the stellar disk may start well inside the solar circle
\citep{lop02}, in which case
the assumption of the constant fraction underestimates the SN Ia rate in the inner Galaxy.

The radial surface distribution of old stellar disk is usually described by a
single exponential profile, but
the radial scale lengths quoted in published papers span an enormous range:
1.8 to 6 kpc (Kent, Dame, \& Fazio 1991 and references therein). Recent near-infrared
studies based on the the {\em COBE}/DIRBE data or the
Two Micron All Sky Survey data appear to yield
a shorter scale length (2.1--2.8 kpc) consistently
(Freudenreich 1998; Ojha 2001; Drimmel \& Spergel 2001 and references therein).
On the other hand, studies on extragalactic SNe show that, although the number of SN
Ia is small (54) to constrain the slope very well,
the distribution could be represented by a single
radial scale length of $R_{25}/5.5$ where $R_{25}$ is the semimajor axes of
isophote having 25.0 mag arcsec$^{-2}$ in blue \citep{van97}.
The Andromeda galaxy (M31), for example, has
$R_{25}=20$~kpc (Trimble 2000), so that the radial scale
length of SN Ia would be $3.6$~kpc.
We adopt 3.0 kpc as the radial scale length of the SN Ia
surface density distribution in the Galaxy,
and assume that the distribution extends all the way to the Galactic center.
The stellar disk may have a central hole where a stellar bar is located \citep{fre98}.
But, since the surface density profile averaged over the
azimuth remains approximately exponential to the center
\citep{fre98}, the assumption of
an axisymmetric exponential distribution is plausible.
Hence, after normalization, the SN rate
per unit area in our model disk is given by
\begin{equation}
\sigmasn(r)=\sigma_0 \exp(-r/3.0~{\rm kpc}) 
\end{equation}
where $\sigma_0 = 5.5 \times 10^{-5}$~kpc$^{-2}$ yr$^{-1}$.
With the above distribution, the fraction exploding inside the central hole
$\fhole=0.34$. Therefore, the fraction of isolated SNe that explodes
within our
model gaseous disk $\fhi=(1-\fhalo)(1-\fhole)=0.44$. The frequency of
\schi SNRs in our model \schi disk, therefore, is
$\Sigma_{\rm SN}\fhi=0.20\times 10^{-2}$~yr$^{-1}$, or one per 500 yrs.

\subsection{Dynamical Evolution of \schi SNRs}

In our model, SNRs evolve in a homogeneous, uniform medium and we are interested in their
late-stage evolution. 
(The effect of embedded clouds will be discussed in Section 4.1.) 
SNRs form a dense, neutral shell through radiative cooling when the
cooling time of the shock-heated material becomes shorter than the age of the SNRs.
Numerical studies showed that the shell
formation occurs `catastrophically', accompanying multiple
shocks and oscillations of shock velocity
\citep[e.g.,][]{fal81, kim97, blo98}.
%(e.g., Falle 1981; Kimoto \& Chenoff 1997; Blondin et al. 1998).
The shell is also dynamically unstable to
develop nonspherical structure \citep{blo98}. But
this violent transition lasts only for a relatively short time, and
the global dynamics of radiative SNRs for most of time
could be described by a spherically-symmetric,
steady model
\citep[e.g.,][]{cox72, che74, cio88}.
We use the results of \cite{cio88} in the following
to describe the SNR evolution. Their model neglects magnetic field and 
thermal conduction. 
Magnetic field inhibits compression to make the shell 
thicker and the outer radius larger. 
Thermal conduction lowers the temperature and increases the density in the 
hot bubble, the effectiveness of which depends on the magnetic field 
configuration. 
These effects, however, are dynamically important only 
at very late times when the shell is sufficiently slowed down. 
For example, the
radius and velocity of the shell in the numerical result of  
\cite{sla92}, which includes
a uniform 5~$\mu$G magnetic field and unimpeded thermal conduction, 
are described well by the formulae in this section until the expansion 
velocity drops to $\sim 50$~\kms. 
For a tangled random magnetic field, the shell will be more 
compressible and the effect of the conduction will be reduced 
\citep[e.g.,][]{mck95}, so that 
the agreement will persist at lower velocities.  
Since the visible \schi SNRs have large 
($\ge 50$~\kms) expansion velocities (Section 2.4), 
the magnetic field and thermal conduction may be neglected 
in our formulation.

We consider the time
when the first element of shocked-heated gas cools to
zero as the onset of the formation of \schi shell. This shell-formation
time $\tsf$ is given by
\begin{equation}
\tsf=3.61\times 10^4 n_0^{-4/7} E_{51}^{3/14}~{\rm yr}, 
\end{equation}
where $E_{51}$ is the SN energy released to the ISM in units of $10^{51}~{\rm ergs}$, and
we assumed that the metallicity is given by solar abundances.
After some finite violent transition period, the shell expands steadily, driving a
radiative shock. \cite{cio88} introduced an ``offset" power law
that can accurately describe the radius and velocity of the shock
during $0.4 \tsf\simlt t \simlt 13\tsf$.
They introduced a parameter ($\tpds\equiv\tsf/e$ where
$e$ is the base of natural logarithm) representing the beginning of the radiative, or
pressure-driven snowplow stage when the dynamics begins to deviate from
the Sedov-Taylor solution due to radiative losses,
and normalized all the variables using this parameter.
Instead of introducing this extra parameter, we slightly modify their
offset power-law (equation 3.32) using $\tsf$ defined in equation (4) to obtain
\begin{equation}
R_s = \rsf \left({11\over 10} {t\over \tsf} -{1 \over 10}\right)^{3/10}, 
\end{equation}
\begin{equation}
v_s = \vsf \left({11\over 10} {t\over \tsf} -{1 \over 10}\right)^{-7/10}, 
\end{equation}
where the radius and velocity of the SNR shock at the shell-formation time are
\begin{equation}
\rsf=20.0 n_0^{-3/7} E_{51}^{2/7}~{\rm pc},
\end{equation}
\begin{equation}
\vsf=179 n_0^{1/7} E_{51}^{-1/14}~{\rm km\, s}^{-1}. 
\end{equation}
We use equations (5) and (6) to describe the evolution of the
radius and velocity of the \schi shell from $\tsf$ to $\sim 13 \tsf$.
The maximum disagreement between the expressions in equation (5) and
the original expression of Cioffi et al. is $<0.1\%$ during this time period.
For $E_{51}=1$ and $n_0=0.14$~cm$^{-3}$, we have $\tsf=1.11\times 10^5~{\rm yr}$,
$\rsf=46.4~{\rm pc}$, and $\vsf=135$~\kms.

The maximum age of \schi SNRs is defined as the
time when the shock velocity drops to about the ambient sound speed ($\sim 10$~\kms)
so that the interior pressure is comparable to the ambient pressure and the shell
breaks up.  Following \cite{cio88}, we
define this merging time $\tmerge$ as the time when the shock
velocity drops to $\beta(=2)$ times the ambient isothermal sounds speed
$c_s$.  Then, using equation (6) and writing the fractions as decimals,
the lifetime of \schi SNRs
$\tau\equiv \tmerge-\tsf$ becomes
\begin{eqnarray}
\tau & = & {10 \over 11} \left[\left(\vsf\over \beta c_s\right)^{10/7} -1\right] \tsf \\
& = & 2.02\times 10^6 (\beta c_{s, 6})^{-1.43}  n_0^{-0.367} E_{51}^{0.316}~{\rm yr}-0.91\tsf, 
\nonumber
\end{eqnarray}
where $c_{s, 6}=c_s/(10^6~{\rm cm~s}^{-1})$.
If $\vsf/\beta c_s\gg 1$, the second term in the second expression may be
dropped. Since the offset power-law for $v_s$
(equation 6) is accurate until $t\simlt 13\tsf$ or $v_s\simgt 0.16\vsf$,
equation (9) becomes less accurate for small ($\simlt 2$) $\beta$.
For $\beta=2$, $c_{s,6}=1$, and $n_0=0.14$~cm$^{-3}$, $\tau \simeq 1.44\times 10^6$~yr.
Note that, since we require the expansion velocity
of the SNR to be considerably greater than
$\beta c_s$ in order to avoid the background confusion, 
the observerable period of \schi SNRs is considerably shorter than $\tau$ (Section 2.4).

In order to compute the detection probability in Section 2.4, we need the mass of the shell as a function
of its velocity. At the time of shell-formation,
the swept-up mass of H atoms using $\rsf$ in equation (7) is
\begin{equation}
\msf=829 n_0^{-2/7} E_{51}^{6/7}~M_{\odot}. 
\end{equation}
During the shell-formation stage, a good fraction ($\sim 50\%$)
of this mass is compressed into a
dense, neutral shell \citep{che74, man74}, and,
after the formation, the subsequently shocked ambient gas is simply added to the shell.
The hot interior gas also continuously
runs into the dense shell to add extra mass, but its contribution
is relatively small.
We assume that the shell appears with a mass of $\delta \msf$
at $\tsf$ and
the newly swept-up mass is simply added to the shell. Then,
from equations (5) and (6), the \schi mass of the shell $M(v_s)$ is given by
\begin{equation}
M(v_s)=\msf \left[ \left( v_s \over \vsf \right)^{-9/7} - (1-\delta) \right],
\end{equation}
where the second term represents the fraction of the mass left in hot interior at the stage of
the shell formation. We adopt $\delta=0.5$ following the numerical results.

\subsection{Fraction of Observable SNRs}

It is convenient to rewrite the fraction of observable \schi SNRs (equation 2) as
an integral over velocity because the detection probability is a function of $v_s$:
\begin{equation}
\fobs={1\over \Sigma_{\rm SN} \fhi \tau}
\int\,\sigmasn(r) \int_{\vmin}^{\vmax} {p(x,y,v_s) \over |dv_s/dt|} dv_s dA 
\end{equation}
where, from equation (6),
\begin{equation}
\left|{dv_s \over dt}\right| = {77 \over 100} {\vsf\over \tsf} \left(v_s\over \vsf\right)^{17/7}.
\end{equation}
The upper limit $\vmax=\vsf$ and the lower limit $\vmin=\beta c_s$. If
$p=1$ in this velocity interval, or equivalently from $\tsf$ to $\tsf+\tau$, then
$\fobs=1$.

The detection probablity $p(x,y,v_s)$ may be expressed as
a product of two probabilities; $\pgb(x,y,v_s)$, the probability of detection that is
limited by the confusion with the Galactic
background \schi emission, and $\ptn(x,y,v_s)$, that the one is
limited by the telescope sensitivity.
Our intention is to compare our results with the observations described in Section 1, and 
we define $\pgb(x,y,v_s)$ so that a \schi SNR is observable 
if the line-of-sight velocity of its shell is 
sufficiently outside the velocity range of the background emission, i.e.,
\begin{displaymath}
\pgb(x,y,v_s)=
\left\{ \begin{array}{r@{\quad\quad}lr@{\quad\quad}lr@{\quad:\quad}l}
1 & {\rm if}~v_{\rm SNR}+v_s \ge v_{\ell, \rm max} + \Delta \vcr \\
  & {\rm or}~v_{\rm SNR}-v_s \le v_{\ell, \rm min} - \Delta \vcr \\
0 & {\rm otherwise,}
\end{array} \right.
\end{displaymath}
where $v_{\rm SNR}$ is the systematic LSR velocity of the SNR and
$v_{\ell, \rm max}$ and $v_{\ell, \rm min}$ are the maximum and minimum LSR velocities
toward $\ell$ permitted by the Galactic rotation. 
We compute the LSR
velocities using a flat rotation curve with $r_\odot = 8.5$ kpc and $v_\odot = 220$~\kms.
The Galactic background \schi emission
extends beyond $v_{\ell, \rm max}$ and $v_{\ell, \rm min}$ because of thermal motions, 
turbulent motions, non-circular motions associated with spiral density waves, etc, and  
$\Delta \vcr$ represents this extra velocity extent of the background emission.
Inspection of large-scale position velocity diagrams suggests that 
$\Delta \vcr$ varies with Galactic longitudes and latitudes.
We adopt $\Delta \vcr=50$~\kms, which is sufficiently large 
in most parts of the Galaxy. 
The dashed lines in Fig. 1 (and Fig. 8) show the boundary 
of the background emission defined in this way. 
It is obvious that the expansion velocity of \schi SNR 
must be greater than 50~\kms\ in order to be visible. When its expansion velocity 
is 50~\kms, the age and radius of the \schi SNR 
$t=3.85\tsf=4.27\times 10^5$~yr and $R_s=1.53\rsf=71.0$~pc from equations 
(5) and (6), respectively.

$\ptn$, the detection probability limited by the telescope sensitivity, depends on
how the SNR couples with
the telescope beam, i.e., whether the SNR is resolved or unresolved by the telescope beam.
We consider two limiting cases:

{\em Model A: Flux-unlimited ($\ptn=1$)}--- This model applies to
an ideal case in which the visibility is not limited by the telescope
sensitivity. The detection is only limited by the confusion due to the Galactic
background emission.

{\em Model B: Flux-limited ($\ptn\le 1$)}--- This model applies to observations
with limited telescope sensitivity. We consider only the case in which the
SNRs are unresolved by the telescope beam. If a source is unresolved, the
21-cm line flux is proportional to the \schi mass of the source and inversely
proportional to the square of the distance.
Therefore, there is a maximum distance $\dmax$ that it is detectable, i.e.,
$\ptn=0$ if $d>\dmax$. The maximum distance is derived below.

Since the \schi 21-cm line emission from the shell is optically thin,
the integrated line intensity (expressed in antenna temperature $T_A$)
is proportional to the total \schi mass $M(v_s)$:
\begin{equation}
\int T_A dv= {3 h c A_{10} \over 32 \pi k m_{\rm H}} {\aeff M(v_s) \over d^2}, 
\end{equation}
where $A_{10}=2.86888\times 10^{-15}$ s$^{-1}$ is the spontaneous transition probability
between the hyperfine structure levels of {\sc Hi}, 
$\aeff$ is the effective area of the telescope,
and the other symbols have their usual meanings. In deriving equation (14),
we assumed that the SNR size is much smaller than the telescope beam.
The SNR size at the shell-formation is $2\rsf=92.8~{\rm pc}=32'\times (d/10~{\rm kpc})^{-1}$, 
and larger at later times. Hence, it appears that most \schi SNRs should have sizes
larger than the beam size ($\sim 30'$) of a
25-m telescope (cf. Fig. 5). However, the excess emission at extreme velocities
 that we  would detect comes from small cap portions of expanding shells, and
$T_A$ from equation (14) might yield an approximate average intensity of the emission.
If the shell is thin and expands
with a uniform velocity $v_s$, then the line profile will be
a rectangular shape with velocity extent
from $-v_s$ to $+v_s$, so that the left-hand side of
equation (14) becomes $T_A\times 2v_s$.
Hence, if
the minimum detectable antenna temperature of the observation is $\Delta T_A$, then
from equations (11) and (14), the maximum distance $\dmax(v_s)$ is given by
\begin{equation}
\dmax (v_s) = {\dsf\over \sqrt\delta} \left(v_s \over \vsf \right)^{-1/2}
\left[ \left( v_s \over \vsf \right)^{-9/7} - (1-\delta)\right]^{1/2}, 
\end{equation}
where $\dsf$ is the maximum distance at the shell-formation stage:
\begin{eqnarray}
\dsf & = & \left[ {3 h c A_{10} \over 32 \pi k m_{\rm H}}{\aeff \over \Delta T_A}
{\delta\msf\over 2\vsf} \right]^{1/2} \\
& = & 1.89 (A_{\rm eff,6}/\Delta T_{A,-1})^{1/2}
\delta^{1/2} n_0^{-3/14} E_{51}^{11/28}~{\rm kpc},\nonumber
\end{eqnarray}
with $A_{\rm eff,6}=\aeff/(10^6 {\rm cm}^2)$ and $\Delta T_{A,-1}=\Delta T_A/(0.1~{\rm K})$.
The probability $\ptn$ in Model B is given by
\begin{displaymath}
\ptn(x,y,v_s)=
\left\{ \begin{array}{r@{\quad\quad}l}
1 & {\rm if}~d\le \dmax(v_s) \\
0 & {\rm otherwise.}
\nonumber\end{array} \right.
\end{displaymath}
For a 25-m telescope with $A_{\rm eff,6}=3.0$ and $\Delta T_{A,-1}=1.0$, 
$\dsf=3.5$~kpc for $n_0=0.14$~cm$^{-3}$ and $\delta=0.5$.

\section{RESULTS}

The basic results are shown in Fig.~2 where we plot the expected surface density
distribution of \schi SNRs in the Galactic plane.  Fig. 2(a) shows the distribution of
all \schi SNRs, i.e., when $\pgb=\ptn=1$.
The total number of \schi
SNRs in our model gaseous
disk is $\Sigma_{\rm SN} \fhi \tau=2850$. This is
44\% of the total ($6480$) isolated SNRs in the Galaxy. The rest
are in the halo (34\% or 2200) and in the central hole (22\% or 1430).
The distribution in Fig. 2(a) is simply
due to the exponential distribution of the SN surface density.
Fig. 2(b) shows the distribution of
observable \schi SNRs in Model A, i.e., when $\ptn=1$.
The total number of observable SNRs is now $267$, or
9.4\% of the total \schi SNRs.
They are concentrated along the loci of
tangential points 
because the systematic velocities of the SNRs in those regions are close to 
either the maximum or the minimum velocities of the Galactic background emission.
Only the receding (approaching) portion of the shell will be detectable 
for the \schi SNRs in the first (fourth) quadrant.
The concentration near $\ell=0^\circ$ is because the minimum (maximum) 
velocity permitted by the Galactic rotation is close to zero in the first (fourth) quadrant. 
For those SNRs, only the approaching (receding) portion of the shell will be detectable 
in the first (fourth) quadrant.
Near $\ell=180^\circ$, both the receding and approaching portions are detectable
(cf. Fig. 4).
Fig. 2(c) is for Model B
when the observation is made with a radio telescope with
an effective area of $3\times 10^6$~cm$^2$, which corresponds to a telescope with a
diameter of 25~m, and $\Delta T_{A,-1}$=1.0.
Now, the SNRs on the far side of the Galaxy are not observable and the
total number is 96, or 3.4\% of the total \schi SNRs.

Fig. 3 (left-hand panel) is the distribution of expansion velocities of \schi SNRs.
The distribution of all \schi SNRs is given by
$dN/dv_s\propto v_s^{-17/7}$ (equation 13). Among these \schi SNRs, the ones
with small
expansion velocities are largely unobservable because of the contamination
due to the Galactic background emission, which is shown by the Model A distribution.
On the other hand, the Model B distribution shows that the SNRs with large expansion
velocities are largely unobservable mainly because of the limited telescope sensitivity.
Fig. 3 (right-hand pannel) is the distribution of
distances to \schi SNRs. The bowl between $d=5$ and 12 kpc in the distribution of
all \schi SNRs is due to the central hole. The rapid drop beyond 7--8 kpc in Model A
distribution is due to the small number of observable SNRs beyond tangential points
in the inner Galaxy (cf. Fig. 2b). The Model B distribution shows that,
with a 25-m telescope, we can detect all the observable \schi SNRs to 4 kpc and most up
to 7 kpc, but no \schi SNRs beyond 9 kpc.

A statistical property that can be directly compared with the
observations is the Galactic longitude distribution of \schi SNRs (Fig. 4).
The distribution of all \schi SNRs peaks
toward $\ell=0^\circ$, but not strongly because of the central hole.
The background contamination
brings down the real distribution by a factor of 5--16 depending on $\ell$ (Model A).
The effect is largest at $\ell=10^\circ$--30$^\circ$ (or 340$^\circ$--350$^\circ$)
and smallest at $\ell=160^\circ$--200$^\circ$.
The telescope sensitivity further decreases the number of SNRs, most notably
between $\ell=330^\circ$ and 30$^\circ$ (Model B), which is understandable from Fig. 2(c).
For Model B, Fig. 4 also shows whether the observable portions of the SNRs are
approaching, receding, or both.
Another statistical property that can be directly compared with the observations
is the angular size distribution (Fig. 5). The distribution
of all \schi SNRs peaks at 40$'$--50$'$. In the Model A distribution,
there is a strong peak at 20--30$'$ and a plateau between $30'$ and 70$'$. Because
of the telescope sensitivity,
there is no observable \schi SNRs smaller than 40$'$ and the distribution peaks at
60$'$--70$'$ (Model B).

\section{DISCUSSION}
\subsection{Effect of Clouds}

The real ISM is pervaded by cold \schi clouds and their effect on our results
should be discussed.
The evolution of SNRs in a cloudy ISM has been studied numerically by  
\cite{cow81}. One of their models (model 6) is for the two-phase ISM and applicable 
to our case. They adopted $n_0=0.2$~cm$^{-3}$ and $E_{51}=0.31$, and assumed that 
the Spitzer's standard \schi clouds \citep{spi78} are randomly distributed with 
a relatively large (7\%) volume filling factor (see below). They simulated the 
evolution until the beginning of radiative phase and showed that 
neither cloud evaporation nor the dynamical 
effects of the clouds affect the evolution of SNRs in a significant way.
Their shell-formation time and radius agree with those 
from equations (4) and (7). They did not derive the fraction $\delta$ of 
mass compressed into neutral shell at the time of shell formation, 
but, since the interior structure is only 
slightly modified, we expect that $\delta$ is not significantly 
different from the homogeneous case, i.e., 0.5. 

In radiative phase, the cloud-shell interaction depends on the ratio of their column 
densities. The column density of the shell of the visible \schi SNRs is  
(3.3--8.8)$\times 10^{18}$~cm$^{-2}$ (cf. equation 11), which 
is less than that of typical clouds ($\sim 1\times 10^{20}$~cm$^{-2}$).
Thus, the collisions with clouds will punch
holes in the shell \citep{mck77, ost88}. 
The hot interior gas, however, will drive a shock 
to reform the missing portion of the shell.
Since the column density for a strong shock to be radiative  
$\nrad\simeq 10^{17.5}(v_s/100~{\rm km\ s}^{-1})^4$ \citep{mck87}, 
the required distance for the reformation of the shell is 
$\nrad/n_0\simeq 0.73(v_s/100~{\rm km\ s}^{-1})^4$~pc. This is much less than 
the mean distance for 
a particular portion of the shell to experience successive collisions 
(see below). Hence, in the two-phase ISM, 
the \schi SNRs will be punched by cold clouds as they expand, but the holes will 
be quickly recovered. The shell loses mass and momentum by the 
collision with clouds.  

The evolution of such radiative SNRs in cloudy medium 
has been analyzed by \cite{ost88}.
The key factor that determines the importance of clouds is
$\exp[-(R_s-\rsf)/\lcl]$ where $\lcl$ is the cloud mean free path. 
For example, the mass in the shell at the time of shell formation 
decreases by this factor due to the collisions with clouds as it expands.
If $(R_s-\rsf)/\lcl \ll 1$, therefore, the `cloud-punching' is not important.
If the clouds are spherical with radius $\acl$ and their
volume filling factor is $\fcl$, then $\lcl=(4/3) \acl /\fcl$.
The filling factor of cold clouds can be inferred from their characteristic
and space-averaged densities.
The two-phase equilibrium yields
the density of cold clouds $\ncl\simeq 40$~cm$^{-3}$, which agrees with observations
\citep{mck95}. The space-averaged density of cold \schi near the Sun is
$\simeq 0.3$~cm$^{-3}$ \citep[e.g.,][]{fer95}, so that $\fcl\simeq 0.3/40=0.75$\%.
The characteristic radius of the clouds inferred from the
\schi column density ($N_{\rm cl}=4\times 10^{20}$~cm$^{-2}$)
of `standard' \schi clouds \citep{spi78} is $\acl=(3/4)N_{\rm cl}/\ncl=2.4$~pc.
Therefore, $\lcl\simeq 430$~pc and $(R_s-\rsf)/\lcl \le 5.7\times 10^{-2}$ for 
visible \schi SNRs. This implies that
the cloud punching effect is not important for the evolution of 
visible \schi SNRs and therefore for their statistics.

Recently \cite{hei03} showed that in two regions they studied the cold ISM is
distributed in huge blobby sheets of thickness $\sim 0.11$ and $\simlt 3.6$~pc
with lengh-to-thickness aspect ratios $\sim 280$ and $\sim 70$, and,
proposed that, if these characteristics are general,
the cold ISM could be organized into a small number of
large sheetlike structures instead of a large number of randomly distributed
small clumps. If the ISM morphology is close to this clumpy sheet model, then
the probability for a SNR to cross the sheets will be small and only a 
portion of the SNR will be affected by the interaction, so that the
the statistics of \schi SNRs will be determined by the nature of the 
intercloud medium.

\subsection{Comparison with Observation}

Our results can be compared with results of KH91 and \cite{koo03b}
introduced in Section 1. In short, KH91 did a sensitive survey of \schi 21 cm emission
lines toward 103 northern Galactic SNRs using the Hat Creek 26-m
telescope ($\aeff\simeq 3.0$), and detected faint extended wing at forbidden
velocities toward 15 SNRs.\footnote{After the KH91 survey in 1991, more SNRs have been
identified in radio continuum and in X-ray, and the number of northern ($\delta\simgt
-38.^\circ 5$) SNRs increased from 103 to 152 (Green 2001), so that 
the number would increase to $15\times (152/103)=22$ by scaling. But this
would not affect the discussion in this paper.} The sensitivity (1.5$\sigma$) of the
survey was $\sim 0.06$~K. \cite{koo03b} searched for similar \schi
features toward 97 southern SNRs and identified another 10 SNRs.
The Southern Galactic Plane Survey data
was obtained with the Parkes 64-m telescope ($A_{\rm eff,6}=18$).
The sensitivity ($1.5\sigma$) of the survey was 0.09--0.14 K.
These 25 (=15+10) SNRs constitute the \schi SNR candidates (see comment in Section 1).
On the other hand, the
expected number from Model B for the parameters of KH91,
e.g., $A_{\rm eff,6}/\Delta T_{A,-1}=3.0/0.6=5$, is $\simeq 120$, and
for those of \cite{koo03b} is $\simeq$190--220. Therefore, the
observed number of \schi SNRs is less than the expected by a factor of $\simgt 5$.
Fig. 6 shows the Galactic longitude distribution of
the \schi SNR candidates. For comparison, we show the expected distributions
and also the distribution of all known Galactic SNRs in the
Green's catalog (Green 2001). 
Most candidates in the first quadrant have an excess emission
at positive velocities, while those in the fourth quadrant at negative velocities.
This seems to be consistent with our expectation (Fig. 4). Fig. 7 shows the angular size
distribution. Most (72\%) Galactic SNRs have sizes smaller than 30$'$, while the average
size of \schi SNR candidates is $46'$. A significant (36\%) fraction of \schi SNR candidates 
has size less than $30'$ whereas none is expected from the model.

The discrepancy between the observed 
and the expected numbers is significant and needs to be explained. 
A plausible explanation is that
the old SNRs are too faint to be identified in radio continuum, so that
the \schi surveys by KH91 and \cite{koo03b}, which were
based on the current (radio continuum-based) SNR catalog, might have missed
most of the SNRs with \schi shells. This thought led us to search for
\schi SNRs {\em based on the \schi data}.
We used the Leiden-Dwingeloo (LD) \schi survey data, which was
obtained by \cite{har97}
using the Dwingeloo 25-m radio telescope (HPBW=36$'$).
The survey covers the Galactic plane ($|b|\simlt 10^\circ$)
between $\ell=0^\circ$ and 260$^\circ$, and
the rms sensitivity of the final data cube is 0.07~K.
We first examined about 130 SNRs in the Green's catalogue and have found
high-velocity features toward $\le 10$ SNRs.
These SNRs
look protruding from their surroundings in ($\ell$, $v$) diagram (cf. Fig. 1).
We then examined individual ($\ell,v$) maps with the naked eye, and
have identified $\sim 70$ similar-looking, forbidden-velocity features
not related to the known SNRs.
Fig. 8 is an example where we can see several such features.
Some of these ``forbidden-velocity (FV) wings" are 
associated with galaxies or 
perhaps HII regions, but most (70\%) of them are not.
Could they be the so-far undiscovered old \schi SNRs? 
We have been looking into these regions in various wavelengths, but 
have not identified their natures yet.
But the distribution of FV wings appears to be very different from
the expected distribution of visible \schi SNRs (Fig. 9): Between
$\ell=0^\circ$ and 70$^\circ$, the number of FV wings is much 
less than
the expected number of \schi SNRs, while, between $\ell=70^\circ$ and
$250^\circ$, it is significantly greater in general.
Possible candidates for such extended
forbidden-velocity \schi wings other than SNRs are stellar wind-blown shells,
neutral stellar winds associated with protostars
\citep{liz88}, high-velocity and/or intermediate clouds, galaxies, etc.
The detailed results are in preparation \citep{kan03}.

If the observed number ($\le 25$) turns out to be close to the
total number of \schi SNRs in the Galaxy, then the discrepancy 
implies that either some of our input parameters are inaccurate or 
some of our basic assumptions are wrong. 
The parameters and assumptions regarding the frequency and 
distribution of SNe, however, are chosen to 
minimize the number of \schi SNRs when they are uncertain, e.g., 
we have assumed that all core-collapse SNe explode in groups and 
only SNe Ia produce isolated SNRs. The total SN Ia rate 
cannot be much lower than what we adopted ($0.45\times 10^{-2}$) if 
we consider that there are at least two SN Ia (Tycho and SN 1006)
during last 2,000 years in the solar neighborhood. 
We have discussed the effect of magnetic fields, thermal conduction, 
and embedded clouds on the evolution of SNRs in Sections 2.3 and 4.1, 
and concluded that they are not dynamically important for the visible
\schi SNRs which are moving fast. 
If the \schi disk is denser and thinner, the number of 
visible \schi SNRs will decrease, but not by a large factor.
For example, if the density of the \schi disk is 
$\sim 0.5$~cm$^{-3}$ and its thickness is $\sim 200$~pc, the number 
decreases by $\sim 40$\%.
One possibility is that 
the interstellar space in the inner Galaxy 
is not filled with WNM, but mostly filled with a very hot,
diffuse ($3.5\times 10^{-3}$~cm$^{-3}$) medium as in the three phase model of \cite{mck77}. 
In such a tenuous medium, the SNRs
develop dense shells when they are very old ($9\times 10^5$~yr, equation 4), at which
time their radius is 230 pc. But this radius is very large, so that
it is possible that the SNRs come into equilibrium with the interstellar pressure
or overlap with other SNRs before dense shell formation, in which case \schi shells
are not formed \citep{mck77}.
Embedded clouds could be accelerated by the SNR shock and also by the
ram pressure, but numerical
simulations by \cite{cow81} showed that they are substantially destroyed by evaporation
before acquiring high velocities.
In the three-phase ISM, the observed \schi SNRs may
be the ones in special environments, e.g., near large atomic or molecular clouds.

The above conclusion on the structure of the ISM in the inner Galaxy
would not be totally changed 
even if the FV wing sources turn out to be \schi SNRs
because, according to Fig. 9, there are much less number of FV wings than
what the model predicts in the inner Galaxy. In the outer Galaxy, however,
the number is significantly greater than the predicted. The model distribution 
can be made to have more \schi SNRs in the outer Galaxy, for example,
by increasing the radial scale length of SNe distribution, but it cannot be 
made to be consistent with the observed FV wing distribution simply by 
changing some of its parameters. 
We postpone further discussion on the FV wings and their significance  
after identifying their natures.

\section{SUMMARY}

We have developed a simple model to estimate the number of old, radiative SNRs visible in
\schi 21-cm emission line in the Galaxy. Old SNRs are expected to be surrounded by rapidly
expanding \schi shells, but the contamination due to the Galactic background \schi emission
severely limits their visibility. The visible SNRs are defined as the ones with a sufficiently
large expansion velocity, so that their line-of-sight velocities are significantly ($\ge 
50$~\kms) outside the velocity range of the Galactic background \schi emission.
Another factor that determines the visibility is the telescope sensitivity, which is
included in our formulation.
Some of the essential features of our model are
(1) the \schi gas in the Galaxy is confined to an axi-symmetric, center-emptied
disk with an inner edge at 3.5 kpc and outer edge at 15 kpc,
(2) the \schi disk is uniform and homogeneous, and the density of the disk is
$0.14$~cm$^{-3}$, (3) the disk is rotating with a constant speed $v_\odot=220$~\kms,
(4) only SNe Ia produce isolated SNRs,
(5) total Galactic SN Ia rate is $0.45\times 10^{-2}$~yr$^{-1}$, but
only 44\% of them explodes within the gaseous disk to produce \schi SNRs
(34\% explodes above and below the disk while 22\% within the central hole), and
(6) the SNe Ia are exponentially distributed with a radial scale length of 3 kpc.
We use the equations of \cite{cio88} in a slightly modified form to
calculate how the radius, velocity, and mass of a SNR shell
vary with time, and, at each position in the disk, we derive the number of
visible SNRs.

According to our result, the number of \schi SNRs in the Galaxy is 2850.
Galactic background contamination limits
the number of visible SNRs to $\simeq 270$, or $\simeq 9$\% of the total \schi SNRs.
They are concentrated along the loci of tangential points and near $\ell=0^\circ$
(Fig. 2b). The telescope sensitivity may prevent observing the ones on
the far side of the Galaxy (Fig. 2c).
We have calculated the expected number of visible \schi SNRs
for realistic observational parameters and compared the results
with observations. There have been two systematic \schi 21-cm emission line studies toward
103 northern and 97 southern SNRs, and faint extended wings at forbidden velocities
have been detected toward 25 SNRs \citep[KH91;][]{koo03b}. The expected number
is much ($\simgt 5$) greater than this.

We consider that our estimate is on the conservative side and
the discrepancy is significant.
A plausible explanation is that previous
observational studies, which were made toward the SNRs identified in radio continuum,
missed most of the \schi SNRs because they are too faint to be visible in radio continuum.
We propose that the faint extended wings at forbidden velocities seen in
\schi 21-cm line surveys (e.g., Fig. 8) could be possible candidates for such
old \schi SNRs. According to our our preliminary result, however, 
their Galactic longitude distribution is quite different from
the expected distribution of visible \schi SNRs, e.g., 
the number of FV wings is considerably less than
the expected in the inner Galaxy. The natures of the 
FV wings need to be identified. 
If the observed number ($\le 25$) turns out to be close to the total
number of \schi SNRs in the Galaxy, then
the discrepancy implies that either some of our input parameters are inaccurate or
some of our basic assumptions are wrong.
A possible explanation is that the interstellar space in the inner Galaxy
is largely filled with a very tenuous gas as in the
three-phase ISM model. In this case,
the observed \schi SNRs are likely the
ones in special environments, e.g., near large atomic or molecular clouds.

\section*{Acknowledgments}

        We wish to thank Chris McKee and Seung Soo Hong for carefully reading
the manuscript and for helpful comments. We thank Carl Heiles and
Jos\'e Franco for helpful discussions. We also thank the anonymous referee 
for comments which improved the presentation of this paper. 
This work has been supported by the Korea Research Foundation
Grant (KRF-2000-015-DP0446).

{}

\clearpage
%\bsp

\begin{figure}
\vspace{20cm}
\includegraphics{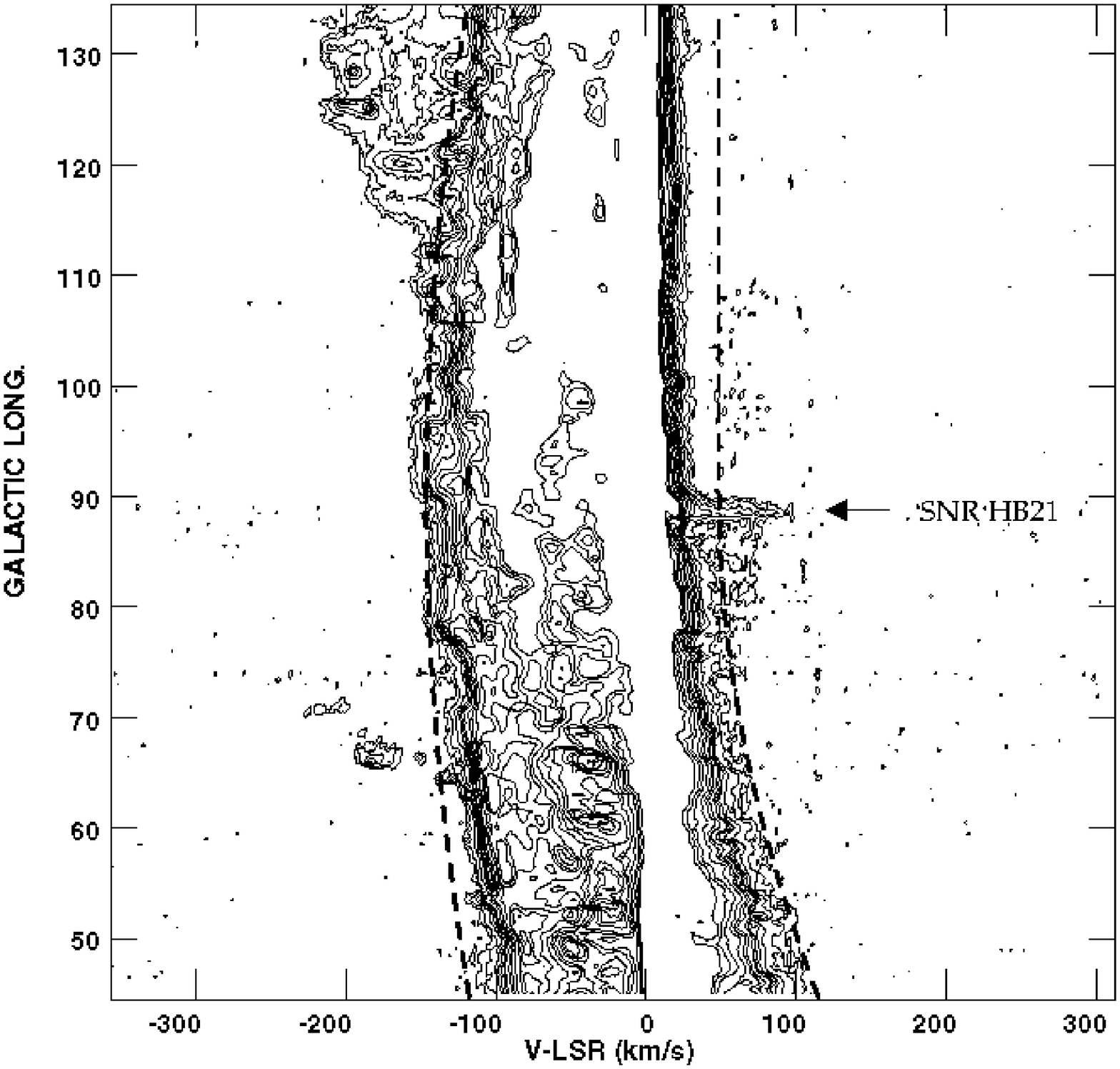}
%%\vspace{256pt}
\caption{
($\ell, v$) diagram of \schi 21-cm emission at $b=5.^\circ 0$. 
Notice the excess emission at the position of the SNR HB 21 ($\ell=89^\circ$) 
between 40 and 100~\kms. 
%The positional coincidence and the forbidden velocity 
%strongly suggest that the excess emission is emitted from the gas accelerated by the SNR
%shock. The systematic velocity of the SNR is $\sim 0$~\kms. 
The dashed lines represent the boundaries of the Galactic background 
emission used in our model (see Section 2.4). 
The map has been made from the Leiden-Dwingeloo \schi survey data \citep{har97}, and 
the contour levels start at 0.1 K and end at 10 K.} 
\end{figure}

\clearpage

\begin{figure*}
\vspace*{20cm}
\includegraphics{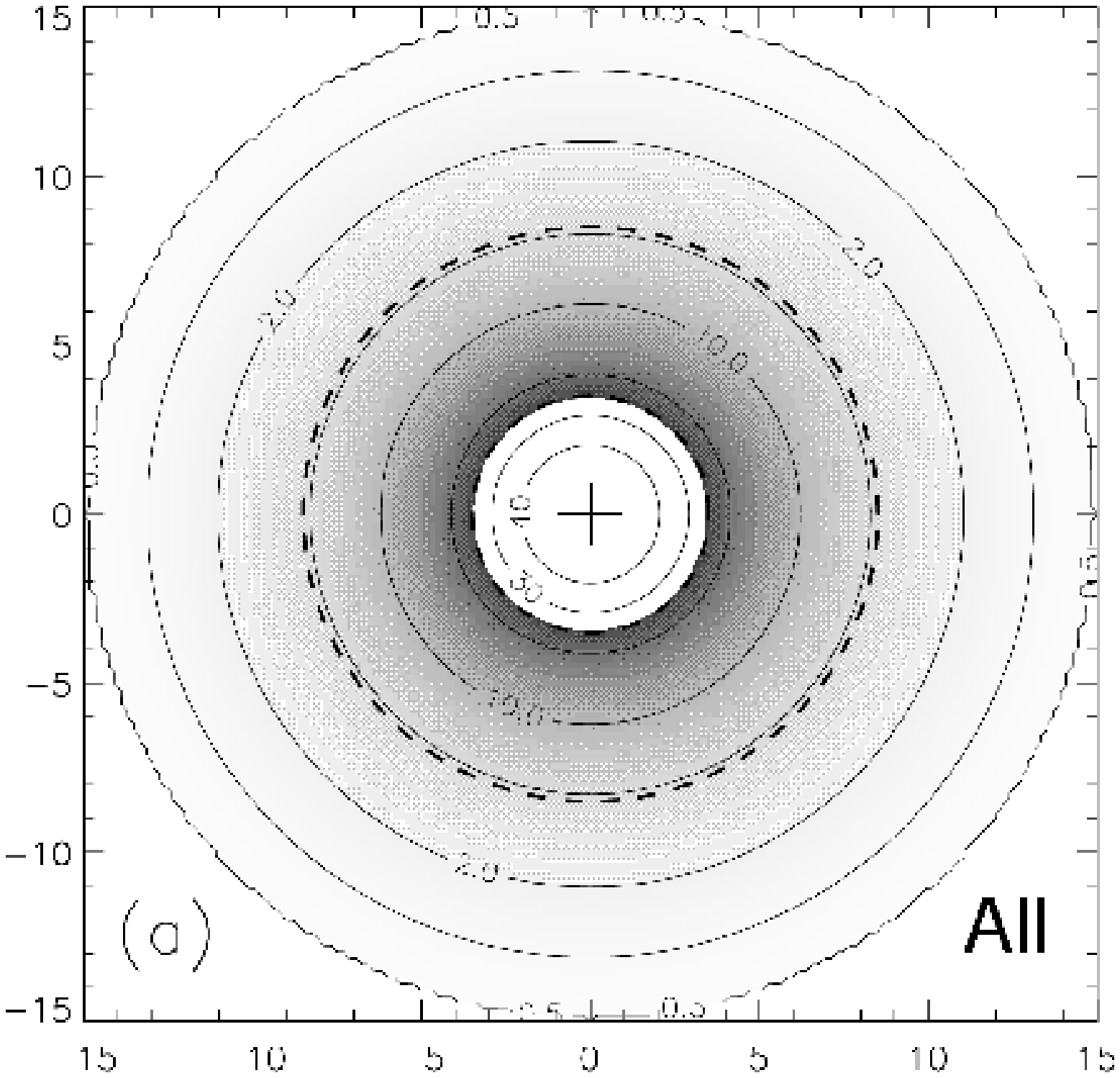}
\includegraphics{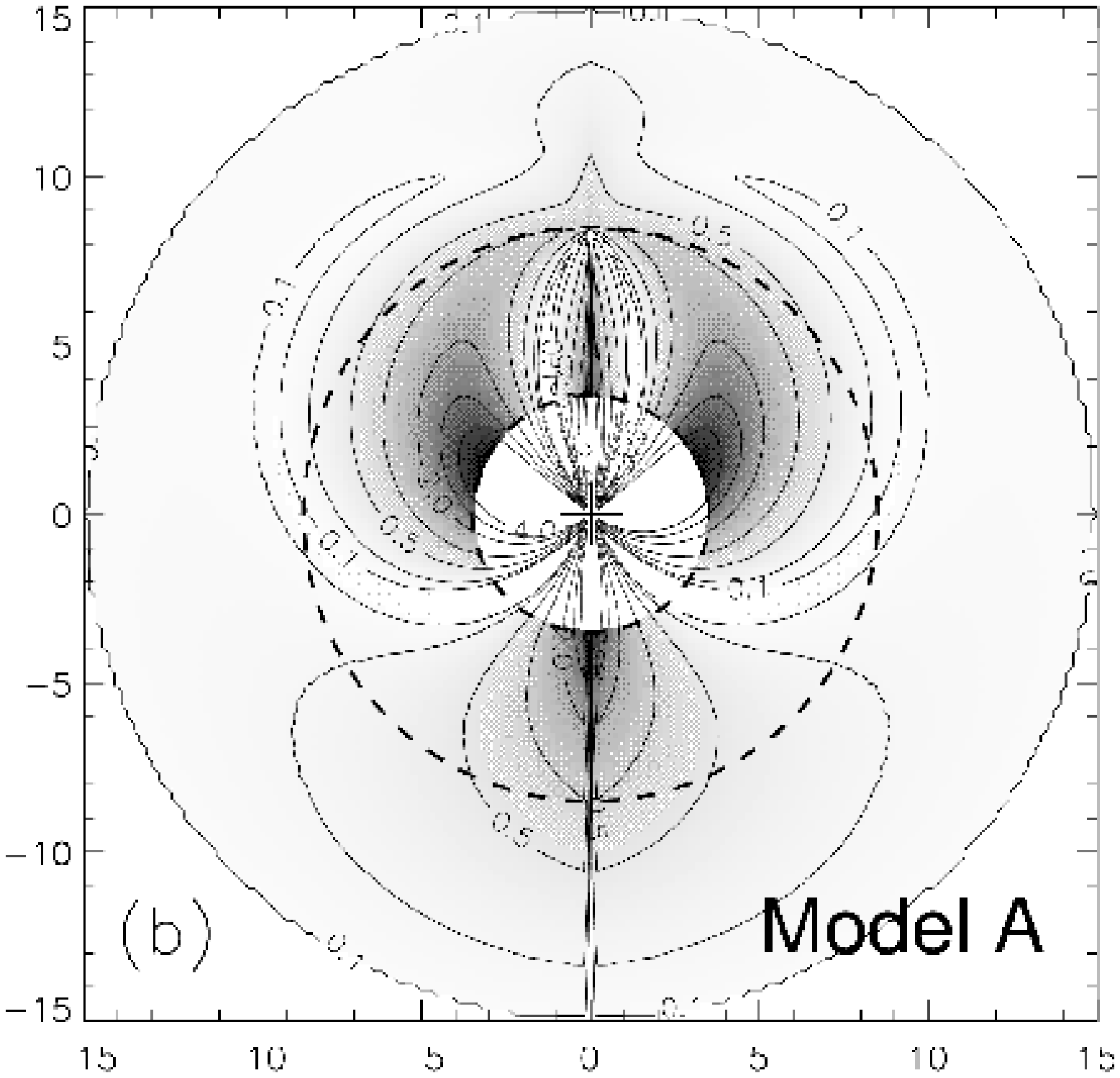}
\includegraphics{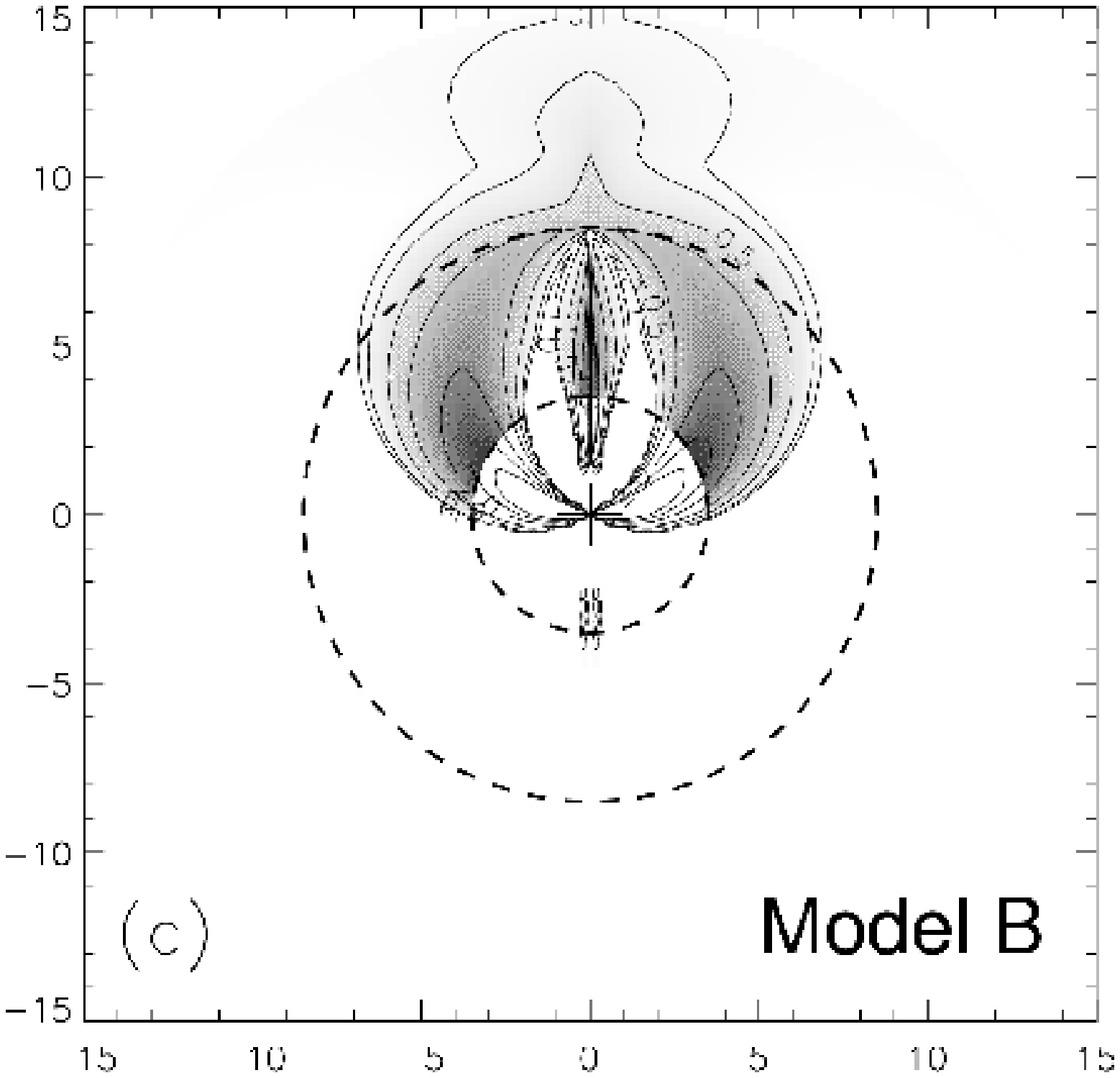}
%%\vspace*{174pt}
\caption{ Expected surface density (kpc$^{-2}$) distribution of 
(a) all \schi SNRs ($\pgb=\ptn=1$), 
(b) observable \schi SNRs in Model A ($\ptn=1$), and 
(c) observable \schi SNRs in Model B ($A_{\rm eff,6}/\Delta T_{A,-1}=3.0$). The inner and 
outer circles marked by dashed lines are at $r=3.5$~kpc and 8.5~kpc, respectively.
In our model, the region inside the inner circle is devoid of the ISM and there the SNe 
explosion do not produce \schi SNRs.
}
\end{figure*}

\clearpage

\begin{figure*}
\vspace*{20cm}
\includegraphics{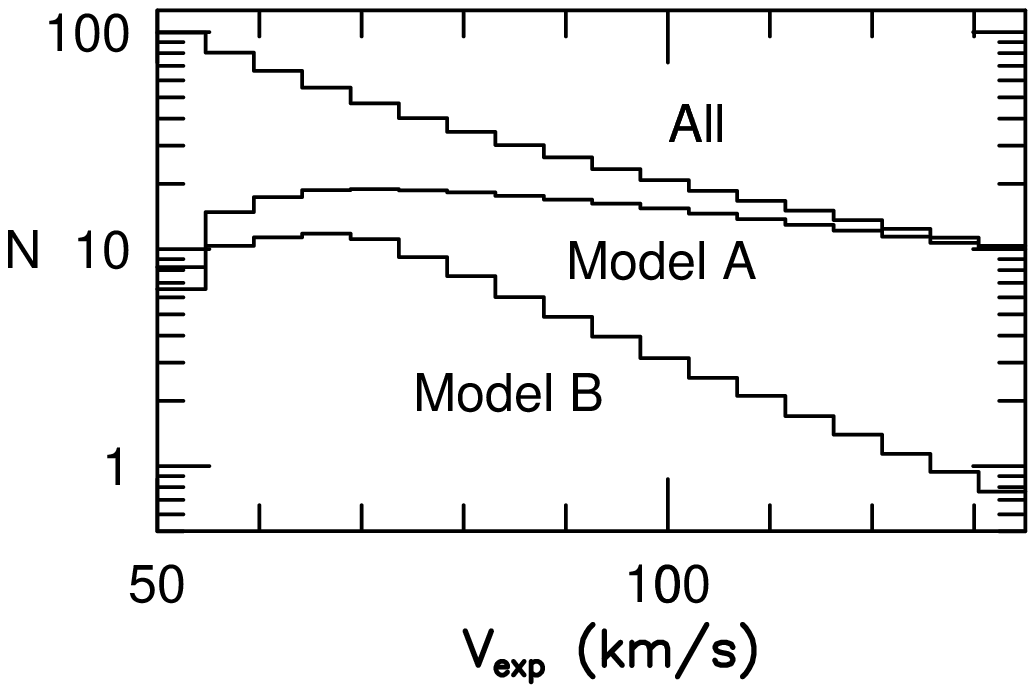}
\includegraphics{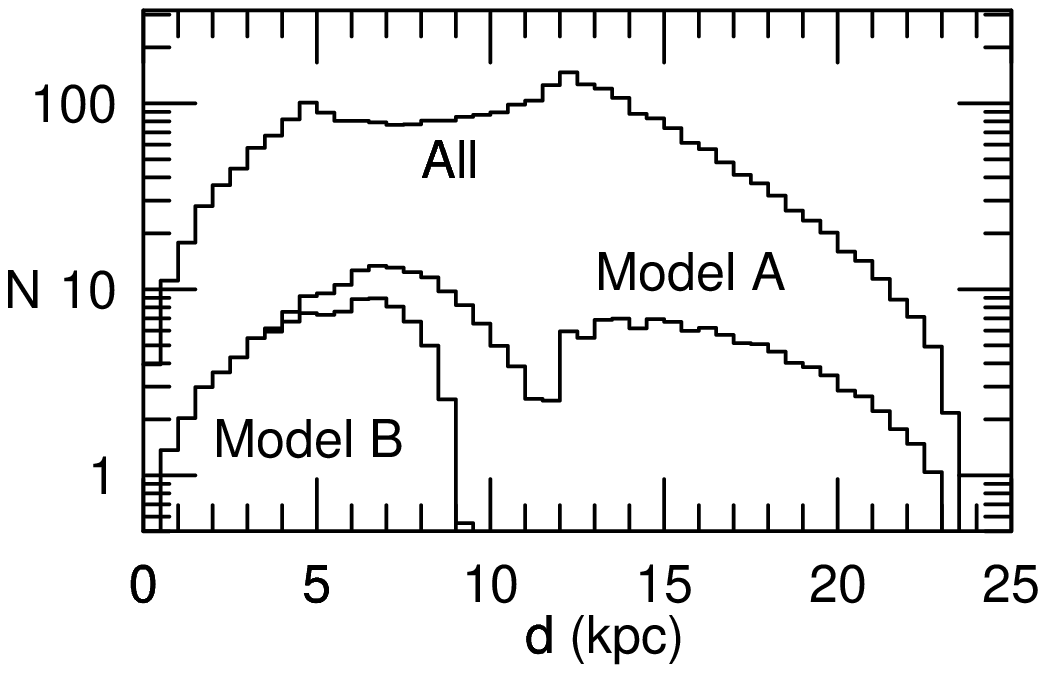}
%%\vspace*{174pt}
\caption[]{
Expected distribution of \schi SNRs in: (left-hand panel) expansion velocities; 
(right-hand panel) distances.
}
\end{figure*}

\clearpage

\begin{figure}
\vspace{20cm}
\includegraphics{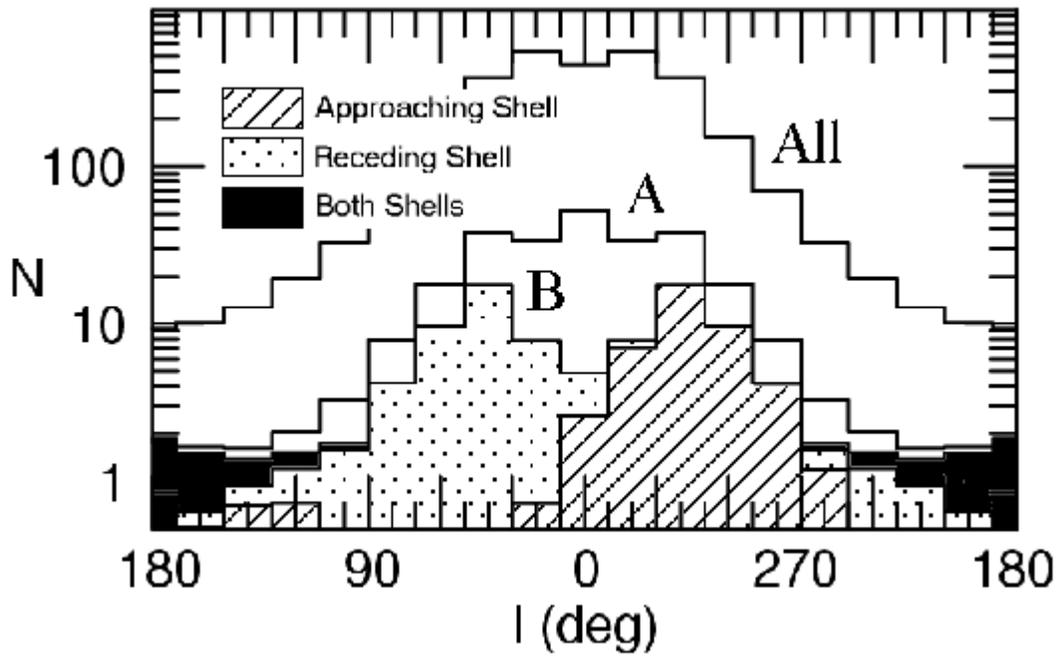}
%%\vspace{174pt}
\caption{ 
Expected Galactic longitude distribution of \schi SNRs. For the distribution in Model B, 
we mark whether the observable portions of the SNRs are approaching, receding, or 
both.}
\end{figure}

\clearpage

\begin{figure}
\vspace{20cm}
\includegraphics{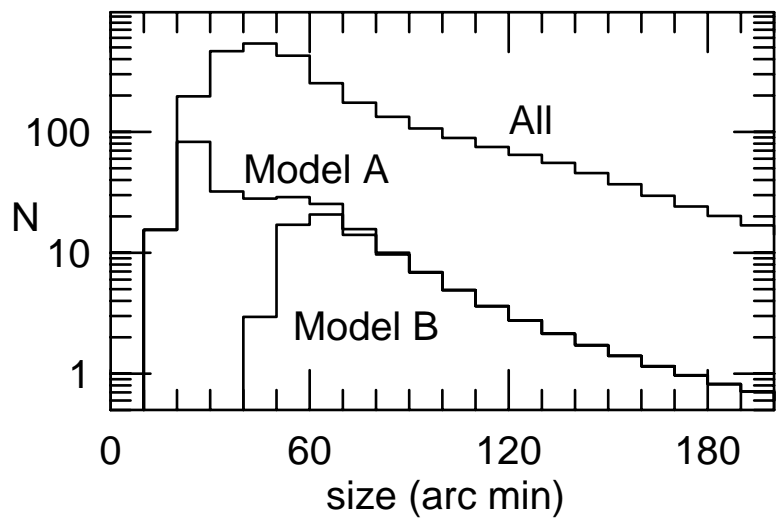}
%%\vspace{174pt}
\caption{ 
Expected angular size distribution of \schi SNRs.}
\end{figure}

\clearpage

\begin{figure}
\vspace{20cm}
\includegraphics{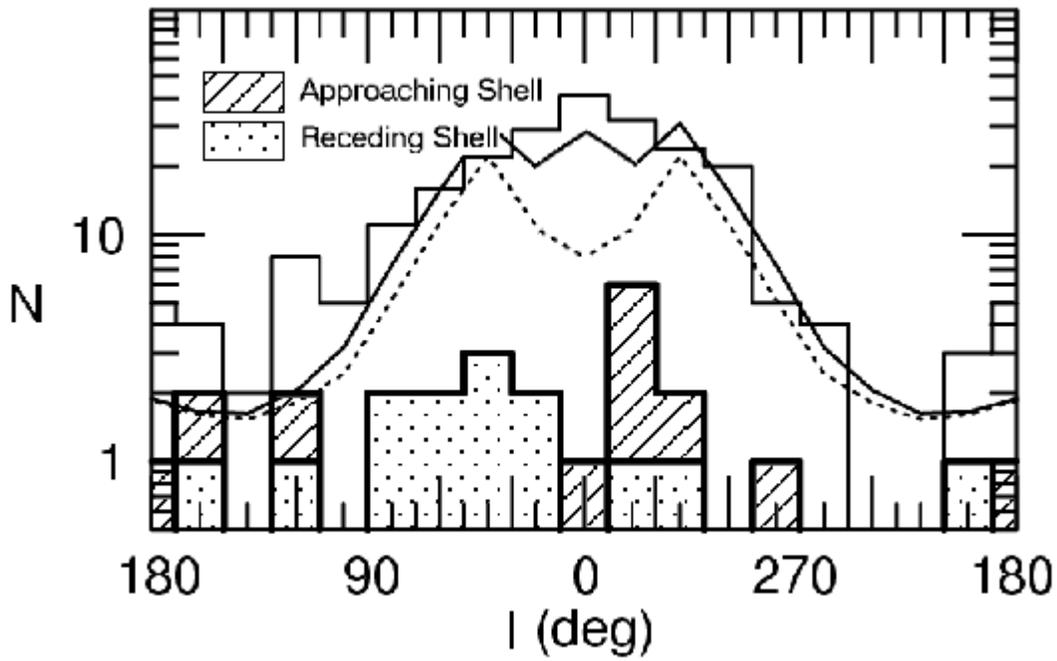}
%%\vspace{174pt}
\caption{
Galactic longitude distribution of SNRs. The distribution of the observed 
\schi SNR candidates is shown by a thick histogram, where we mark whether their 
approaching or receding portions are detected. 
For comparison, the expected distributions 
for the parameters of KH91 and Koo et al. (2003) are 
shown by solid and dotted lines, respectively. The distribution of all known
231 SNRs is also shown as a thin histogram. 
}
\end{figure}

\clearpage

\begin{figure}
\vspace{20cm}
\includegraphics{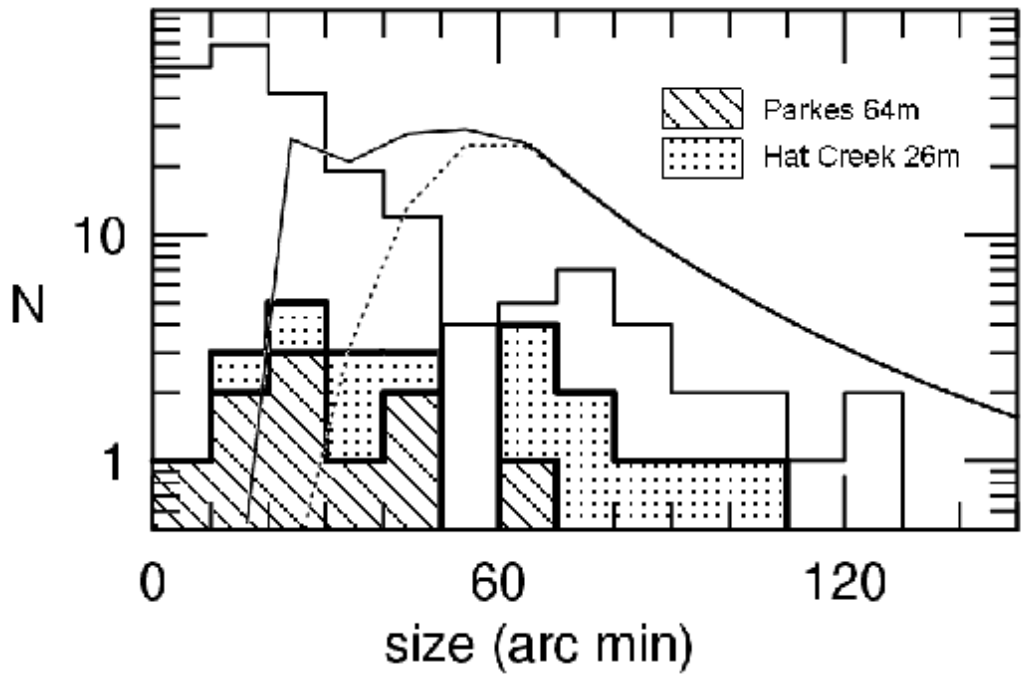}
%%\vspace{174pt}
\caption{
Angular size distribution of SNRs. The meanings of curves and histograms are same as 
those of Fig. 6. For the distribution of the \schi SNR candidates, 
we mark the contributions from the northern (Hat Creek 26m) and 
southern (Parkes 64m) sky surveys, separately.
}
\end{figure}

\clearpage

\begin{figure}
\vspace{20cm}
\includegraphics{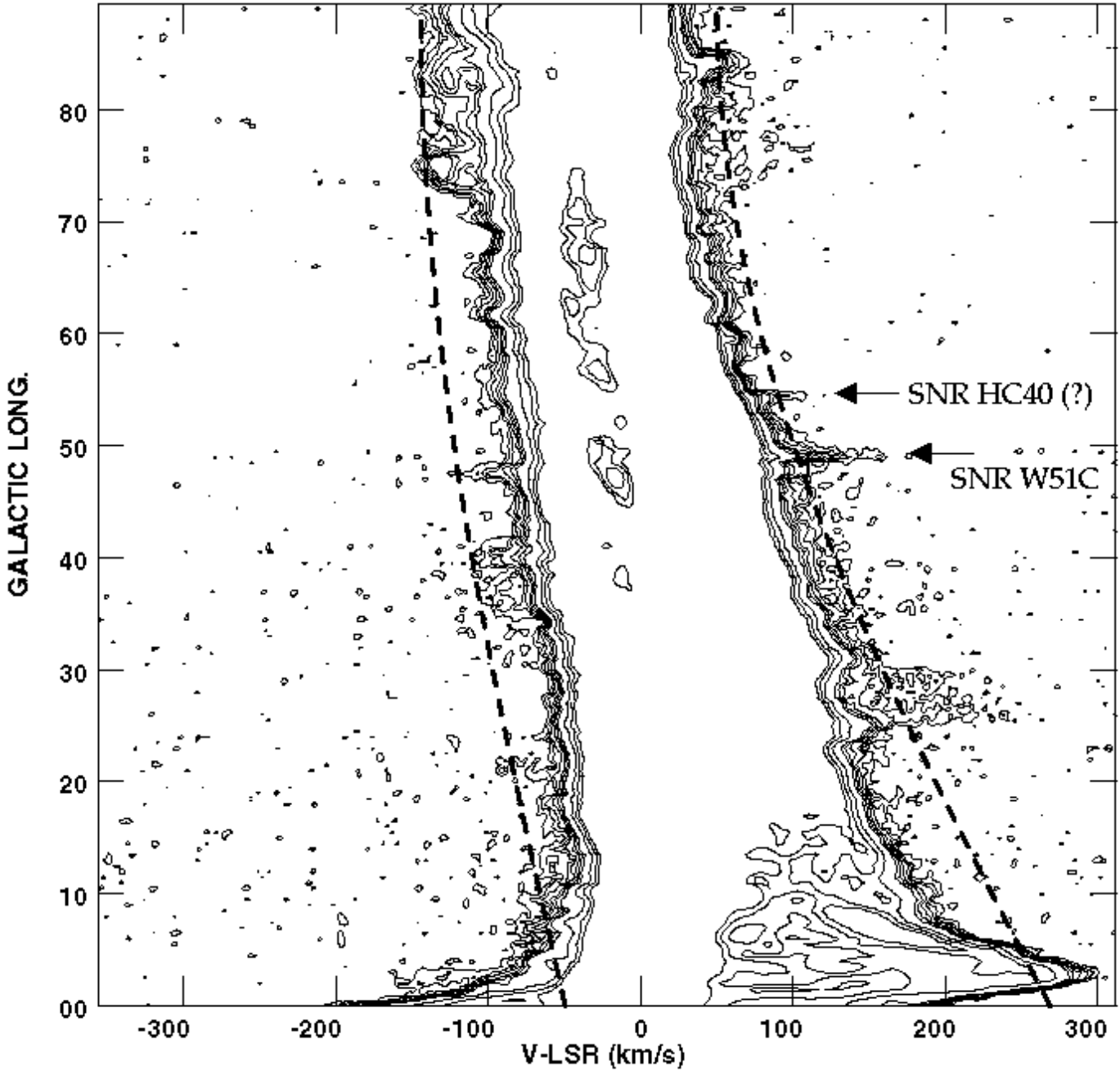}
%%\vspace{256pt}
\caption{
($\ell, v$) diagram of \schi 21-cm emission at $b=-0.^\circ 5$. 
Contour levels start at 0.1~K and end at 10~K.
Note the spatially-confined, faint extended wings at 
forbidden velocities (FV wings). 
Some of them are associated with SNRs, e.g., 
the one between $\vlsr=80$ and 140~\kms\ near 
$\ell=49^\circ$ (SNR W51C) and also, possibly, the one 
between $\vlsr=80$ and 110~\kms\ near 
$\ell=54^\circ$ (SNR HC 40). But the others are  
not related to the known SNRs.
The dashed lines are the same as those in Fig. 1.
}
\end{figure}

\clearpage

\begin{figure}
\vspace{20cm}
\includegraphics{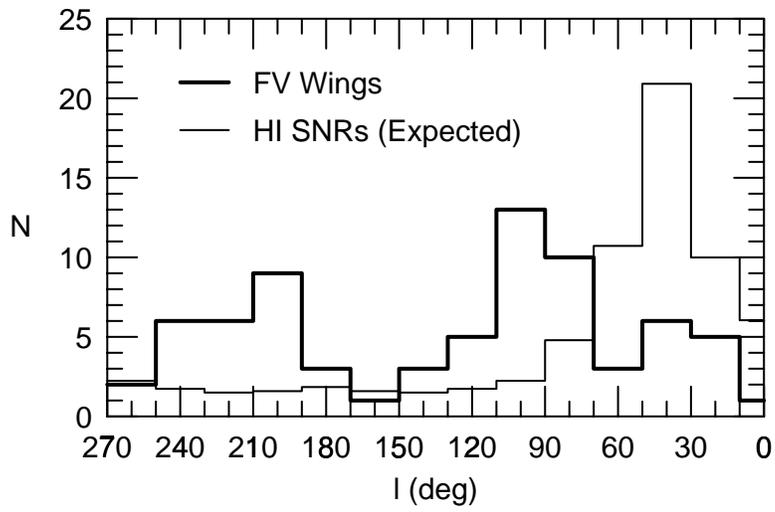}
\caption{
Galactic longitude distribution of FV wings compared to the 
expected distribution of \schi SNRs.
}
\end{figure}

\label{lastpage}
\end{document}